\documentclass[prb,aps,twocolumn,floatfix,showpacs]{revtex4}
\usepackage{graphicx}
\usepackage{amssymb}

\begin{document}
 
\title{Interplay of anisotropy and frustration:
triple transitions in a triangular-lattice antiferromagnet}

\author{P.-\'E. Melchy and M. E. Zhitomirsky}
\affiliation{
CEA, INAC, Service de Physique Statistique, Magn\'etisme et Supraconductivit\'e,
F-38054 Grenoble, France
}
\date{\today}

\begin{abstract}
The classical Heisenberg antiferromagnet on a triangular lattice 
with the single-ion anisotropy of the easy-axis type is studied theoretically.
The phase diagram in an external magnetic field is constructed 
from the mean-field analysis. 
Three successive Berezinskii-Kosterlitz-Thouless transitions are found
by Monte Carlo simulations in zero field. 
Two upper transitions are related to the breaking of the discrete 
$\mathbb{Z}_6$-symmetry, while the lowest transition
is associated with a quasi-long-range ordering of transverse components.
The intermediate collinear phase between the first and second transition
is the critical phase predicted by  J. V. Jos\'e {\it et al}.\ 
[Phys.\ Rev.\ B {\bf 16}, 1217 (1977)].
\end{abstract}
 
\maketitle

\section{Introduction}

Frustrated magnetic systems have been a stimulating research topic over
several decades. Their diverse properties, highly degenerate 
ground states, non-collinear ordering, novel phase transitions,\cite{Diep2005} 
offer a playground to investigate fundamental  physical questions 
going far beyond magnetism itself. One of the specific 
subjects in this field is the interplay of geometrical frustration
and magnetic anisotropies. The prominent example is 
provided by the rare-earth pyrochlore materials with Ising-type
magnetic moments. Contrary to naive expectations, these magnetic systems
remain non-frustrated for an antiferromagnetic nearest-neighbor coupling,
but develop highly frustrated spin-ice states
for the case of a ferromagnetic exchange
between spins.\cite{Harris97,Moessner98,Ramirez99} 

In the present work we investigate the nearest-neighbor
Heisenberg antiferromagnet on a triangular lattice with 
the single-ion anisotropy of the easy-axis type:
\begin{equation}
{\cal H} = J \sum_{\langle ij\rangle} {\bf S}_i\cdot {\bf S}_j - D\sum_i (S^z_i)^2
\label{Hamilt}
\end{equation}
Such a Hamiltonian is believed to describe quasi two-dimensional 
(2D) magnetic materials ${\rm VCl_2}$ (Ref.~\onlinecite{Kadowaki87}) 
and ${\rm LiCrO_2}$ (Ref.~\onlinecite{Kadowaki95}). 
A similar model with the $XXZ$
anisotropy has been previously studied by a number of
authors.\cite{Miyashita85,Sheng92,Stephan00}
In real magnetic materials with $S>\frac{1}{2}$ the single-ion anisotropy
being the first-order relativistic effect
is usually more significant than the anisotropic exchange,
which is generally of the second-order in 
the spin-orbital coupling. \cite{Yosida96}
Besides, as we shall see later, the two types
of anisotropy lead to different
sequences of finite-temperature phase transitions.

Ordered states of the anisotropic triangular antiferromagnet (\ref{Hamilt})
are characterized by a nonzero static magnetization:
\begin{equation}
\langle {\bf S}_i \rangle = 
{\bf l}_1 \cos ({\bf Q}\cdot{\bf r}_i) + {\bf l}_2 \sin ({\bf Q}\cdot{\bf r}_i) + {\bf m} \ .
\label{OP}
\end{equation}
with the ordering wave vector ${\bf Q}=(4 \pi/3, 0)$.
At zero temperature the Heisenberg triangular-lattice antiferromagnet
orders in a three-sublattice $120^\circ$ spin structure. 
Such a noncollinear magnetic ordering is described by a pair of orthogonal 
antiferromagnetic vectors: ${\bf l}_1 \perp{\bf l}_2$, $|{\bf l}_1|=|{\bf l}_2|$, and
${\bf m}\equiv 0$.
In accordance with the Mermin-Wagner theorem there is no symmetry breaking
transition at any finite temperature. Still a weak topological 
transition related to proliferation of $\mathbb{Z}_2$-vortices 
may occur for this model at
$T/J\sim 0.3$.\cite{Kawamura84,Wintel94,Southern95,Caffarel01,Kawamura07}
For the easy-plane anisotropy, $D<0$ in Eq.~(\ref{Hamilt}), the spin plane of
the ordered $120^\circ$ structure is fixed to the $x$--$y$ plane.
In this case two finite temperature transitions take place: the Ising-type
transition related to the chiral
symmetry breaking and the Berezinskii-Kosterlitz-Thouless (BKT) transition associated with
the vortex-antivortex unbinding.\cite{Capriotti98}

\begin{figure}[t]
\centerline{
\includegraphics[width=0.75\columnwidth]{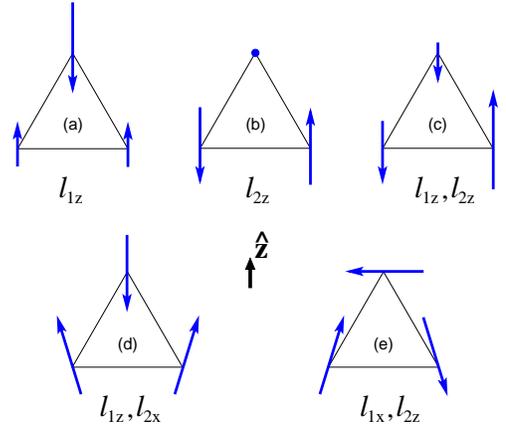}}
\caption{(Color online) Possible three-sublattice planar configurations 
of the easy-axis triangular antiferromagnet.
The direction of the easy-axis is shown by ${\bf \hat{z}}$.
The nonzero components of the order parameter~(\ref{OP}) are
indicated for each configuration.}
\label{config}
\end{figure}

The easy-axis anisotropy, $D>0$, orients the spin plane perpendicular to the
$x$--$y$ crystallographic plane and simultaneously distorts the spin structure.
Finding directions and magnitudes of ${\bf l}_1$ and ${\bf l}_2$
becomes a nontrivial problem in this case. Possible spin structures
corresponding to the ordering wave vector ${\bf Q}$ are presented in Fig.~\ref{config}.
They have been obtained by a symmetry analysis
and are confirmed by the mean-field calculations described in the next section.
Some of these states, Figs.~\ref{config}(a), (c), and (d), have a finite uniform magnetization
$\bf m$ along $\hat{\bf z}$, which is, however, a secondary order parameter and not indicated for 
that reason in the Figure.

In order to elucidate symmetries of different phases, we note that a simple translation 
$\hat{T}_{\bf a}$ (${\bf r}_i \rightarrow {\bf r}_i+{\bf a}$)
transforms the antiferromagnetic order
parameter according to
\begin{equation}
\hat{T}_{\bf a}\bigl[({\bf l}_{1} + i {\bf l}_{2})\bigr] = 
({\bf l}_{1} + i {\bf l}_{2})\;
e^{-i {\bf Q}\cdot{\bf a}} \ ,
\label{groupAction}
\end{equation}
where the phase factor can take only three different values:
${\bf Q}\cdot {\bf a} = 0,\pm 2\pi/3$.
Hence, besides the group $S_1$ of continuous rotations about the $\hat{\bf z}$-axis
the magnetic structure has an inherent discrete symmetry $\mathbb{Z}_3$.
Such an additional symmetry corresponds to permutations of three sublattices.
In zero magnetic field the time-reversal symmetry implies invariance with respect
to ${\bf l}_i \rightarrow -{\bf l}_i$, which enlarges $\mathbb{Z}_3$ to $\mathbb{Z}_6$. 
The total symmetry group is, therefore, 
\begin{equation}
G = S_1 \otimes \mathbb{Z}_6 \ ,
\end{equation}
see also a similar discussion in Ref.~\onlinecite{Sheng92}.
The collinear phases shown in Figs.~\ref{config}(a)-(c)
preserve the axial symmetry $S_1$
but break in different ways the discrete symmetry group $\mathbb{Z}_6$.
In terms of the order parameter angle $\varphi$ defined by
\begin{equation}
l_{1z} = l \cos \varphi\ ,\qquad  l_{2z} = l\sin \varphi \ ,
\label{ls}
\end{equation}
the state in  Fig.~\ref{config}(a) corresponds to
commensurate values  $\varphi = 2k\pi /6$ with an integer $k$,
whereas the configuration in Fig.~\ref{config}(b) has 
$\varphi = (2k+1)\pi/6$.
The third type of a collinear state
is described by an arbitrary angle $\varphi$ and is shown
schematically in Fig.~\ref{config}(c).
In such a state the phase $\varphi$ remains unlocked
and  the sine and cosine harmonic (\ref{ls})
coexist with an arbitrary ratio.

For large enough values of 
$D/J> d_{c}=1.5$ the magnetic anisotropy induces a 
highly degenerate collinear Ising state at zero temperature. 
Quantum fluctuations can lead, then, to interesting 
zero- and finite-temperature phases.\cite{Damle07,Sen09}
Here, we investigate an antiferromagnet with  a moderate-strength anisotropy 
$0<D/J<d_c$, which is frequently found among experimental
systems, and consider the finite-temperature properties of the model (\ref{Hamilt}). 
For simplicity, we neglect quantum effects and study the classical
spin model. 

Layered easy-axis triangular antiferromagnets with a significant 
interplane coupling exhibit two second-order
transitions with an intermediate collinear $l_{1}$-phase 
shown in Fig.~\ref{config}(a).\cite{Plumer88} In contrast, we
show in the present work that a purely 2D system (\ref{Hamilt}) shows 
three consecutive BKT-type transitions. 
In the first part, Sec.~II,  we investigate
the mean-field phase diagram 
of the model (\ref{Hamilt}) at zero and at finite magnetic fields.
The mean-field behavior is expected to be 
realized in layered triangular antiferromagnets
with weak interplane coupling.
The Monte Carlo (MC) simulations  and the analysis 
of the zero-field behavior  of the model
(\ref{Hamilt}) are presented in the second part of our study, Sec.~III.

\section{Mean-field theory}
\label{RSMFT}

Let us begin with the mean-field analysis of possible finite-temperature
phases of the model (\ref{Hamilt}). Specifically, we use the
real-space approach,\cite{Bak80,Suzuki83,Cepas04,Enjalran04}
generalizing  the previously established technique to systems with the single-ion
anisotropy.
The two standard steps of the mean-field approximation include
(i) decoupling the spin-spin interaction according to
\begin{equation}
{\bf S}_i \cdot {\bf S}_j \approx
{\bf S}_i\cdot \langle {\bf S}_j \rangle +
\langle{\bf S}_i \rangle \cdot{\bf S}_j - \langle{\bf S}_i\rangle \cdot \langle{\bf S}_j\rangle \ ,
\end{equation}
with  $\langle{\bf S}_i\rangle$ being the thermal average of an $i^{\rm th}$ magnetic moment
and (ii) rewriting  ${\cal H}$ 
as a sum of single-site Hamiltonians
\begin{eqnarray}
{\cal H}_{\rm MF} & = & \sum_i \big[-D (S^z_i)^2-{\bf h}_i\cdot {\bf S}_i\big]
 - J \sum_{\langle ij\rangle} \langle{\bf S}_i\rangle \cdot \langle{\bf S}_j\rangle \nonumber \\
& & {\rm with}\ \ \ {\bf h}_i = {\bf H} - J \sum_{\rm n.n.} \, \langle{\bf S}_j\rangle \ ,
\label{Hamiltf}
\end{eqnarray}
where we have also added a Zeeman magnetic field to  Eq.~(\ref{Hamilt}).
Due to the presence of the single-ion term in ${\cal H}_{\rm MF}$,
the local magnetization $\langle {\bf S}_i\rangle$ has to be decomposed 
into components, which are  transverse  and parallel to the anisotropy axis:
\begin{equation}
\langle {\bf S}_i \rangle= \langle S^z_{i}\rangle \, {\bf {\hat z}}
+ \langle S^\perp_i\rangle \,\frac{[{\bf h}_{i} - h_{i}^z{\bf {\hat z}}]}{h_{i}^{\perp}} \ .
\end{equation}
Performing integration with respect to $x = S^z_i = \cos\theta_i$ in the expression
for the partition function we obtain the following mean-field 
equations for static magnetic moments:
\begin{eqnarray}
\langle S_i^\perp \rangle &=&
\frac{1}{2 Z_{i}} \int_{-1}^1\mathrm{d} x\,\sqrt{1-x^2}\,e^{Dx^2/T}\,e^{h_{i}^z x/T}
 I_1(y_{i})\ ,
\nonumber  \\
\langle S_{i}^z\rangle & = & \frac{1}{2Z_{i}} \int_{-1}^1 \mathrm{d} x\, x \, e^{Dx^2/T}\, e^{h_{i}^z x/T}\; 
I_0  (y_{i})\ , \label{selfconsistent_class} \\ 
Z_{i} &=& \frac{1}{2} \int_{-1}^1 \mathrm{d} x\; e^{Dx^2/T}\: e^{h_{i}^z x/T} 
I_0  (y_{i})\ , \nonumber 
\end{eqnarray}
where $y_{i} = h_{i}^{\perp}\sqrt{1-x^2}/T$ and $I_n(z)$ is the modified Bessel function of 
the $n$-th order:
$$
I_n(z) = \frac{1}{\pi} \int_0^\pi \mathrm{d} \alpha \,   e^{z\cos\alpha}\, \cos^n\!\alpha\ .
$$

\begin{figure}[t]
\centerline{
\includegraphics[width=0.75\columnwidth]{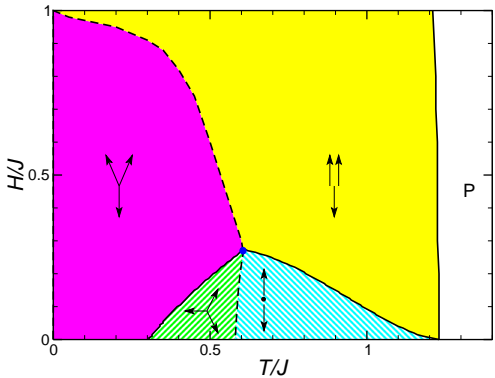}}
\caption{(Color online) 
The low-field part of the mean-field phase diagram with
${\bf H}\parallel \hat{\bf z}$ for a representative value $D/J=1$.
Spin configurations for each phase are schematically indicated by arrows as in
Fig.~\ref{config}. 
Solid and dashed lines correspond to first- and second-order transitions, 
respectively.
}
\label{phasesMF}
\end{figure}

The system of integral equations (\ref{selfconsistent_class}) 
together with the self-consistency condition given by Eq.~(\ref{Hamiltf}) 
is solved iteratively on finite 
lattices of $N=L\times L$ spins, with periodic boundary conditions.  
Once convergence is achieved, various physical quantities are 
calculated  including the free-energy
\begin{equation}
{\cal F}_{\rm MF} = 
 - J \sum_{\langle ij\rangle} \langle{\bf S}_i\rangle \cdot 
\langle{\bf S}_j\rangle -T \sum_i \ln Z_i \ ,
\end{equation}
the internal energy $E_{\rm MF} = \langle {\cal H}_{\rm MF}\rangle$,
and the antiferromagnetic order parameters. By explicit calculations
for clusters with $3\leq L \leq 12$ at all temperatures and weak magnetic fields
we have verified stability of the three-sublattice
structure with ${\bf Q} = (4\pi/3,0)$. After that 
a more detailed investigation of the $H$--$T$ phase diagram has been performed
with the three-sublattice ansatz. Precise location of phase
boundaries in Fig.~\ref{phasesMF}
has been determined from  temperature and field scans 
for the  antiferromagnetic order parameters indicated in Fig.~\ref{config} 
as well as for the uniform magnetization. The behavior of the specific heat 
has been also used to independently verify these results.

At the upper transition $T_{c1}\simeq 1.2J$ in zero magnetic field 
only $z$-components of magnetic moments become ordered.
In accordance with the $\mathbb{Z}_6$ symmetry 
selection between various collinear structures is determined
by the following invariant in the Landau free-energy:
\begin{equation}
A_6 \bigl[ (l_{1}^z+il_{2}^z)^6 + \textrm{c.\,c.}\bigr] \ .
\label{sixthorder}
\end{equation}
For negative $A_6<0$  the pure $l_1$-state, Fig.~\ref{config}(a), is energetically favored, 
while  $A_6>0$ corresponds to the $l_2$-state, Fig.~\ref{config}(b). 
We have verified the positive  sign of $A_6$ in our case by a direct analytical expansion of 
Eqs.~(\ref{selfconsistent_class}).
Our numerical results also confirm  that the $l_2$-state is stable below $T_{c1}$.
Such a partially ordered phase has a vanishing moment on one of the antiferromagnetic
sublattices. A similar phase has been discussed in relation to the  intriguing
phase diagram of ${\rm Gd_2Ti_2O_7}$.\cite{Stewart04} Here, we provide an
example, where a partially ordered phase is realized at the mean-field level
in a simple spin model.

The second transition at 
$T_{c2}\simeq 0.6J$ is related to the breaking of the rotational symmetry 
about $\hat{\bf z}$-axis. Below $T_{c2}$ the third 
previously disordered magnetic sublattice becomes ordered
with moments oriented  within the $x$--$y$ plane.
Simultaneously, moments of the other two sublattices
start deviating from $\hat{\bf z}$-axis leading to  
a distorted triangular structure shown in Fig.~\ref{config}(e).
This distorted spin structure is characterized by ${\bf l}_2 \parallel \hat{\bf z}$
and ${\bf l}_1 \perp {\bf l}_2$.
When temperature is further decreased the coefficient $A_6$ in the effective
anisotropy  term changes  sign at $T_{c3}\simeq 0.3$
and one finds a first-order transition into another distorted triangular structure
shown in Fig.~\ref{config}(d) with
${\bf l}_1 \parallel \hat{\bf z}$.

Note, that the related model with the exchange anisotropy 
\cite{Miyashita85,Sheng92} has $A_6=0$ in the mean-field approximation,
which  leads to an additional continuous degeneracy.
As a result, only two finite-temperature transitions are found in this case: 
from the paramagnetic state to a degenerate collinear configuration 
shown in Fig.~\ref{config}(c) and then to 
a degenerate distorted $120^{\circ}$ configuration.\cite{Miyashita85,Stephan00}
Sheng and Henley \cite{Sheng92} have discussed how different types
of fluctuations, thermal, quantum, or random dilution,
can induce a finite $A_6$. For the model with the single-ion
anisotropy one finds a different interesting possibility:
the sign of the anisotropic term changes upon lowering temperature.

The two phases in Figs.~\ref{config}(a) and (d) have a nonvanishing  
total magnetization $m^z$. The coupling between ferro- and antiferromagnetic
components is determined by the term
\begin{equation}
m^z (l_{1}^z+il_{2}^z)^3 + \textrm{c.\,c.} \ ,
\label{couplingFMAFM}
\end{equation}
which is invariant under $\mathbb{Z}_3$ transformations (\ref{groupAction}).
In zero magnetic field this yields $m^z \sim (T_c-T)^{3/2}$ 
for states with ${\bf l}_1 \parallel \hat{\bf z}$.
In contrast, states in Fig.~\ref{config}(b) and \ref{config}(e)
with ${\bf l}_2 \parallel \hat{\bf z}$ have vanishing $m^z$.
This difference is important to understand the finite-field
behavior, see Fig.~\ref{phasesMF}.
Magnetic field applied parallel to the $\hat{{\bf z}}$-axis favors spin structures
with a finite magnetization and stabilizes states with $l_1^z \neq 0$,
which is why the two intermediate low-field phases are no longer
pure $l_{2}^z$ states.  This feature is emphasized by 
hatches in Fig.~\ref{phasesMF}.
The collinear-noncollinear transitions are of the second order, whereas
all other transition lines are of the first order. In the case 
of the transition from the paramagnetic state in external magnetic field
the first-order nature of the transition follows from the presence of the cubic
invariant (\ref{couplingFMAFM}), while in other cases the above
conclusion is a consequence of the group-subgroup relation.
The transition lines intersect at a multicritical point $(T^*,H^*) = (0.6J,0.25J)$.

The mean-field phases and the structure of the phase diagram at fields larger than
$H^*$ are similar to the Heisenberg triangular antiferromagnet \cite{Kawamura85}
 so we do not go into further details. 
We have also checked other moderate values of $D/J < 1.5$
and found precisely the same structure of stable phases
with triple transitions in zero magnetic field.
As we shall see in the next section, the true thermodynamic phases
determined by Monte Carlo simulations of the model (\ref{Hamilt}) differ
from the mean-field solutions, which is often the case in 2D.
Still, the mean-field picture is expected to be qualitatively correct for 3D
layered triangular antiferromagnets. 
By including a ferro- or antiferromagnetic interlayer coupling $J'$
in the mean-field equations (\ref{Hamiltf}) and 
(\ref{selfconsistent_class}) we have verified 
that the predicted sequence of finite-temperature transitions  remains valid 
up to $|J'/J| \sim 0.6$. For larger values of $|J'/J|$ we find a double transition
with an intermediate $l_1$ collinear phase similar to the previously
studied case of very strong $J'$.\cite{Plumer88}

\section{Monte Carlo simulation}
\label{MCsection}

\begin{figure}[t]
\centerline{
\includegraphics[width=0.9\columnwidth]{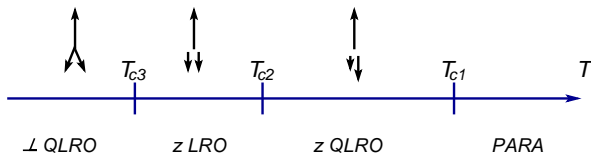}}
\caption{
Schematic zero-field phase diagram of 
the two-dimensional triangular antiferromagnet with easy-axis 
single-ion anisotropy. The arrow labeling of phases is the same as in Fig.~1.
}
\label{MCdiagram}
\end{figure}

In uniaxial magnetic systems, transverse
and longitudinal spin components order at different temperatures
as they belong to different irreducible representations.
For the triangular antiferromagnet with the easy-axis anisotropy,
the highest transition should be related to the sole breaking of
$\mathbb{Z}_6$ symmetry. Such a discrete symmetry breaking may lead
to a phase with a true long-range ordering at low temperatures even
in 2D. The case of a 2D system with the general $\mathbb{Z}_p$ symmetry
has been considered in the seminal work of Jos\'e and co-workers.\cite{JKKN77}
The precise nature and sequence of finite-temperature transitions
depend on the number $p$ of ``clock states.''  Jos\'e \emph{et al.}\
have predicted two BKT-type transitions for $p = 6$. A massive
phase with a true long-range order appears below the lower transition
at $T_{c2}$, while at intermediate  temperatures $T_{c2}<T<T_{c1}$ a gapless
phase with an algebraic quasi long-range order is realized.
In our case the massive phase is represented by one of the 
states in Figs.~\ref{config}(a) and (b), while the gapless phase
correspond to a state in Fig.~\ref{config}(c) with
a power law decay of spin-spin correlations:
\begin{equation}
\langle S_i^z S_j^z \rangle \sim \frac{\cos({\bf Q}\cdot {\bf r}_{ij})}{r_{ij}^\eta} \ .
\end{equation}
The critical exponent $\eta$ continuously varies from $\eta_1=1/4$
at $T=T_{c1}$ to $\eta_2=1/9$ at $T=T_{c2}$.
The subsequent BKT transition related to the appearance of quasi-long-range
order in the transverse components
is expected to occur at an independent transition temperature
$T_{c3} < T_{c1}$.
The expected sequence of finite-temperature phases is schematically shown
in Fig.~\ref{MCdiagram} with  three BKT-type transitions.
A similar suggestion was made before for the triangular
antiferromagnet with the exchange anisotropy, \cite{Sheng92}
though no supporting numerical results were presented.

\begin{figure}[t]
\centerline{
\includegraphics[width=0.9\columnwidth]{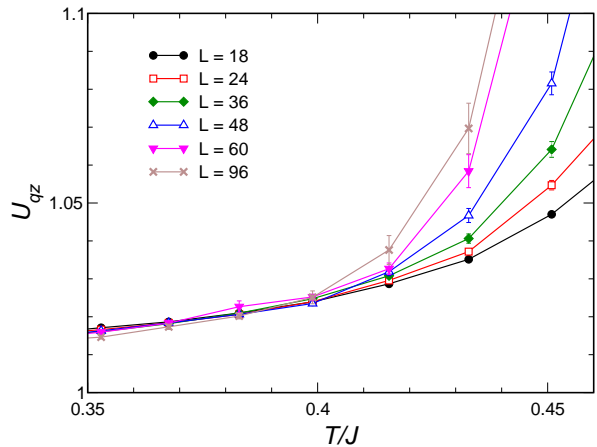}}
\caption{(Color online) 
Temperature dependence of the 
Binder cumulant for the antiferromagnetic order parameter $m_{\bf Q}^z$ 
for different cluster sizes. }
\label{Uqz}
\end{figure}

\begin{figure}[b]
\centerline{
\includegraphics[width=0.9\columnwidth]{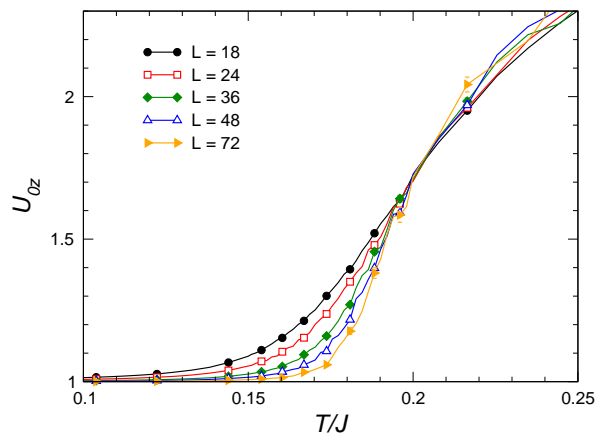}}
\caption{(Color online) 
Temperature dependence of the 
Binder cumulant for the uniform magnetization $m_{z}$ 
for different cluster sizes. }
\label{U0z}
\end{figure}

To verify the outlined scenario in our case 
we have performed Monte Carlo simulations
of the model (\ref{Hamilt}) in zero magnetic field 
for the same value of the anisotropy parameter $D/J=1$ as in Sec.~II.
Rhombic lattice clusters with periodic boundary conditions
and with $N=L^2$ sites, $L=18 - 96$, have been studied using the
standard Metropolis algorithm. Restricted motion of spins was
implemented  at low temperatures to keep the acceptance rate around
$50 \%$. In order to improve further the performance of the MC algorithm,
we have added a few microcanonical over-relaxation steps.\cite{Creutz87,Kanki05}
For models without the single-ion term an over-relaxation move
consists in a random rotation of a given spin about the local magnetic
field. Such a step would not conserve the single-ion energy in (\ref{Hamilt}).
We choose, therefore, to reflect  a spin with respect to the plane
${\bf n}$--${\bf h}$, where ${\bf n}$ is the anisotropy axis and ${\bf h}$
is the local field. In total $2 \cdot 10^6$ hybrid MC steps were used at
each temperature and results were further averaged over
20 different cooling runs, which both reduces measurement noise
and provides an unbiased estimate of the statistical errors.

\begin{figure}[t]
\centerline{
\includegraphics[width=0.9\columnwidth]{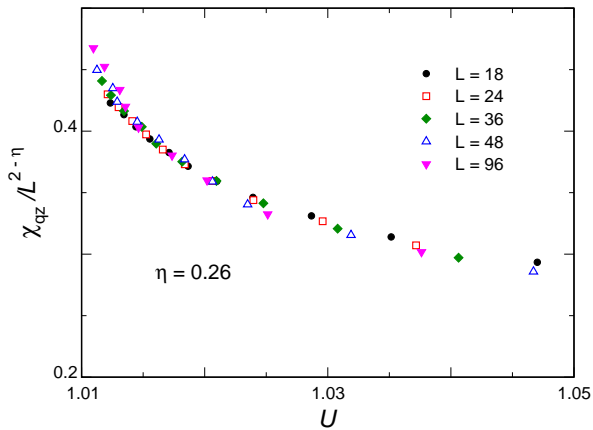}}
\caption{(Color online) 
Scaling plot for the normalized susceptibility versus the Binder cumulant
in the vicinity of the upper transition $T_{c1}$.
The indicated value of the exponent $\eta$ is used to achieve the best 
collapse of data from different clusters.
}
\label{eta1}
\end{figure}

\begin{figure}[b]
\centerline{\includegraphics[width=0.9\columnwidth]{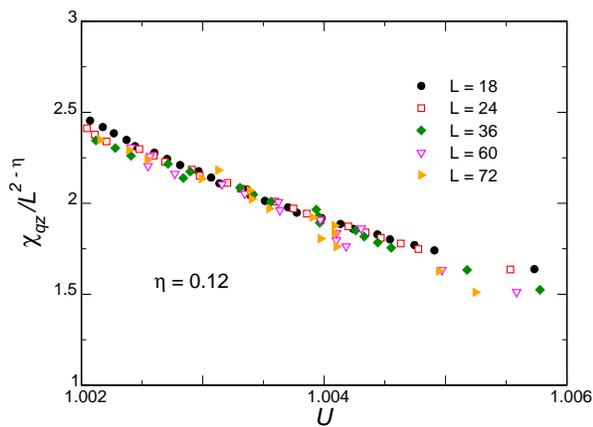}}
\caption{(Color online) 
Scaling plot for the normalized susceptibility versus the Binder cumulant
in the vicinity of the second transition $T_{c2}$.
The indicated value of the exponent $\eta$ is used to achieve the best 
collapse of data from different clusters.
}
\label{eta2}
\end{figure}
 
The standard technique to locate a BKT transition
is to measure the spin stiffness, \cite{Ohta79,Teitel83,Weber88,Chaikin00} 
which jumps from
zero to the universal value $\rho_s = 2T_{\rm BKT}/\pi$. 
However, in the case of an underlying discrete  symmetry definition
of $\rho_s$ becomes problematic. Therefore, we initially focus
on the behavior of the Binder cumulant
$U_{A} = \langle A^4 \rangle / \langle A^2\rangle^2$,
where $A$ is the appropriate order parameter given by Eq.~(\ref{mq2}) below. 
When correlations of the considered order parameter are
critical the value of the Binder cumulant becomes size-independent.
As a result, the curves $U_L(T)$ measured for different cluster sizes $L$ cross 
at the same point for a second-order transition, whereas for 
a BKT transition they merge once the correlation length is infinite. \cite{Loison99}

At every temperature we have separately measured 
even powers of different components of the order parameter
\begin{equation}
(m_{{\bf q}}^{\alpha})^2 = \frac{1}{N^2} \sum_{i,j} \langle S_i^\alpha S_j^\alpha\rangle 
\, e^{i {\bf q} ({\bf r}_i - {\bf r}_j)} 
\label{mq2}
\end{equation}
for $\alpha = z$ and $x,y$ and for ${\bf q}= {\bf Q},0$.
Numerical results for  $z$-components 
are presented in Figs.~\ref{Uqz} and \ref{U0z},
which allow to locate approximately $T_{c1}/J \sim 0.4$ 
and $T_{c2}/J \sim 0.2$. 
For the second transition we use for illustration  the uniform
magnetization  $m^z$ instead of $m_{\bf Q}^z$. Nonzero values of $m^z$ 
unambiguously establish $l_1$-state in Fig.~\ref{config}(a) as the 
low-temperature state with the broken $\mathbb{Z}_6$ symmetry. In addition,
this choice yields less noisy results. Still, statistical errors are significant
and the precise location of the transition point
is difficult with this method.

\begin{figure}[t]
\centerline{
\includegraphics[width=0.9\columnwidth]{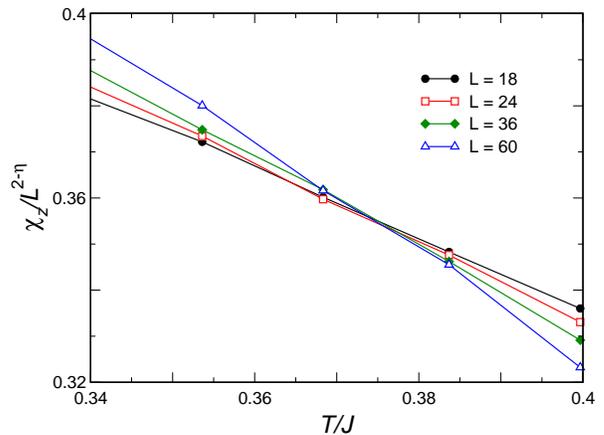}}
\caption{(Color online) 
Temperature dependence of the normalized
susceptibility with $\eta = 1/4$ 
for different cluster sizes 
in the vicinity of the upper transition. The common crossing point yields
$T_{c1}/J \approx 0.377$.
}
\label{Tc1}
\end{figure}
\begin{figure}[b]
\centerline{
\includegraphics[width=0.9\columnwidth]{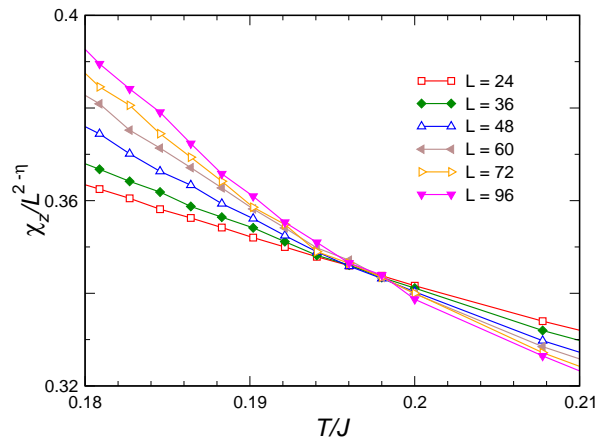}}
\caption{(Color online) 
Temperature dependence of the normalized
susceptibility with $\eta = 1/9$ 
for different cluster sizes 
in the vicinity of the lower transition. The common crossing point yields
$T_{c2}/J \approx 0.198$.
}
\label{Tc2}
\end{figure}

The renormalization group prediction \cite{JKKN77} 
for the exponent $\eta$ in the vicinity of the  two transitions can be, however,
tested without precise knowledge of the corresponding $T_c$. \cite{Loison99}
In the critical regime the general scaling law \cite{Barber83} reads as
\begin{equation}
U_L(T) = f(L/\xi) \quad
\mathrm{and} \quad \chi = L^{2-\eta} g(L/\xi) \ ,
\label{scaling}
\end{equation}
where $\chi = L^2 m^2_{{\bf q}\alpha}/T $ is the generalized susceptibility, 
and $\xi$ is the correlation length.
Hence, the plot of  $\chi/L^{2-\eta}$ against $U_L$ 
for the correct value of $\eta$ should exhibit a collapse of numerical data for 
different cluster sizes onto
a single curve. Figures \ref{eta1} and \ref{eta2} show the best fits
around $T_{c1}$ and $T_{c2}$ respectively, which yield  $\eta_1 = 0.26 \pm 0.01$ and 
$\eta_2 = 0.12 \pm 0.01$. The obtained values are in a very good agreement with the 
prediction $\eta_1 = 1/4$ and $\eta_2 = 1/9$.  \cite{JKKN77}

Once the value of the critical exponent $\eta$ is precisely
established, one can use it to accurately estimate 
the transition temperature from the finite-size scaling of 
susceptibility (\ref{scaling}).  \cite{Cuccoli95,Wysin05}
The curves $\chi/L^{2-\eta}$ for different cluster sizes shown 
in Figs.~\ref{Tc1} and \ref{Tc2} exhibit very tight crossing 
points giving us the following estimates for the transition
temperatures:
$T_{c1}/J = 0.377 \pm 0.001$ and $T_{c2}/J = 0.198 \pm 0.001$.

\begin{figure}[t]
\centerline{ 
\includegraphics[width=0.9\columnwidth]{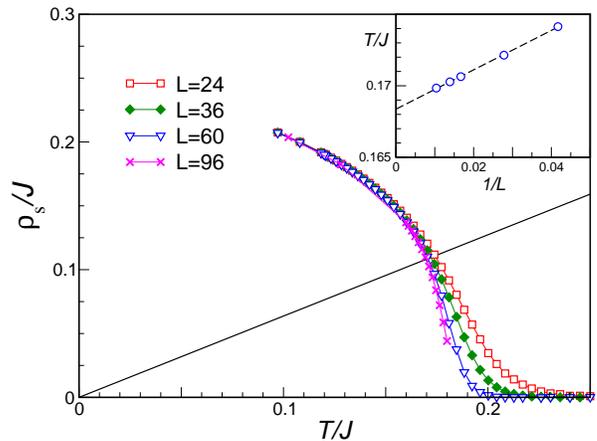}}
\caption{(Color online) The spin stiffness of the easy-axis triangular 
antiferromagnet for different cluster size. Intersection points with 
the line  $\rho_s/T = 2/\pi$ are used to locate
the transition temperature. The inset presents the interpolation
of the crossing  points to the thermodynamic limit.
}
\label{stiff}
\end{figure}

The third BKT transition, which corresponds to a quasi-long-range ordering
of transverse components, occurs at  $T_{c3}<T_{c2}$.
To precisely locate  $T_{c3}$ we measure  the spin stiffness. 
The spin stiffness $\rho_s$ is defined as a general elasticity
coefficient in response to a weak nonuniform twist of spins 
$\delta \phi^\alpha({\bf r})$ performed about a certain direction 
$\alpha$ in spin space.
Generally, the spin stiffness is a fourth-rank tensor with the first pair 
of indexes running over the spin components and the second pair 
spanning over the gradient components in real space. 
In our case it is sufficient to consider only twists about 
the $\hat{\bf z}$-axis in the spin space, while all directions 
in the lattice plane are equivalent 
due to the six-fold rotational symmetry. This leaves us a single parameter:
\begin{equation}
\delta F = \frac{\rho_s }{2} \int \mathrm{d}^2 r \bigl[\nabla \phi^z({\bf r})\bigr]^2  \ .
\end{equation}
Choosing a twist with a uniform gradient 
along an arbitrary direction $\hat{\bf e}$ in the lattice plane, one
obtains in spherical coordinates 
\begin{eqnarray}
{\bf S}_{i} \cdot {\bf S}_{j} & = & \cos \theta_{i} \cos \theta_{j} + 
\sin \theta_{i} \sin \theta_{j} 
\cos ({\tilde \varphi_{i}} - {\tilde \varphi_{j}}) \nonumber \\
&& {\rm with} \ \ \
{\tilde \varphi_{i}} = \varphi_{i} + \delta\phi\, {\bf {\hat e}} \cdot {\bf r}_{i} \ .
\end{eqnarray}
Calculating the change of the free-energy  up to the second order 
in a small $\delta\phi$ and normalizing  result per unit area one obtains
\cite{Ohta79,Teitel83,Weber88} 
\begin{eqnarray}
\rho_s & = &  -\frac{J}{N\sqrt{3}}\, \sum_{\langle i,j \rangle} 
\, \langle (S_{i}^xS_{j}^x + S_{i}^yS_{j}^y) \rangle \\
& & \mbox{} + \frac{2J^2}{NT\sqrt{3}} \Bigl\langle \Bigl\{ \sum_{\langle i,j \rangle} 
(S_{i}^xS_{j}^y\! -\! S_{i}^yS_{j}^x) [\hat{\bf  e}\cdot({\bf r}_{i} - {\bf r}_{j})]  \Bigr\}^2 
\Bigr\rangle  \ .
\nonumber 
\end{eqnarray}
The first term in the above equation has been averaged over $\hat{\bf e}=\hat{\bf x}$ and 
$\hat{\bf y}$ directions. Numerical results from our MC simulations
are presented in Fig.~\ref{stiff}.
We determine crossing points of $\rho^L_s(T)$ with
the straight line  $\rho_s = 2T/\pi$ for each cluster size $L$
and extrapolate them  to $L\rightarrow\infty$
according to $T_{\rm cross}(L) = T_{c3} + a/L$.
This yields the BKT transition at $T_{c3}/J = 0.168 \pm 0.001$ as illustrated in the inset of  Fig.~\ref{stiff}.
We have also determined the critical exponent $\eta_3 =0.28\pm 0.03$,
which coincides within the error bars  with the BKT value
$\eta=1/4$.

\section{Summary}

We have studied a simple model of the Heisenberg
triangular-lattice antiferromagnet with the single-ion anisotropy of the easy-axis type.
Despite its simplicity such a 2D spin model exhibits 
a  sequence of three BKT-type transitions
illustrating  nontrivial physical effects
which appear due to the competition between magnetic anisotropy
and geometrical frustration. The Monte Carlo simulations
yield for $D=J$: $T_{c1}/J= 0.377$, $T_{c2}/J= 0.198$, and $T_{c3}/J= 0.168$.
The two upper transitions correspond to the breaking of the discrete
$\mathbb{Z}_6$ symmetry, whereas the lowest one is the standard
topological transition related to the proliferation
of $XY$ vortices.  At $T_{c2}<T<T_{c1}$
the longitudinal spin correlations have a power law decay with distance 
with a continuously varying exponent $\eta$.

A remaining question is the fate of the intermediate critical phase
at finite magnetic fields. An external field applied parallel to the
anisotropy axis reduces the discrete symmetry from 
$\mathbb{Z}_6$ to $\mathbb{Z}_3$.
According to Jos\'e {\it et al.},\cite{JKKN77} the $p=3$ clock model
has no critical phase but exhibits instead a single transition into 
a normally ordered state, {\it e.g.}, the phase in Fig.~\ref{config}(a). 
It would be interesting to verify numerically the nature of this phase 
transition, which may be a critical one belonging to the three-state 
Potts model universality class \cite{JKKN77} or be of the first order 
due to a presence of the cubic term (\ref{couplingFMAFM}). 

The mean-field calculations find the partially ordered collinear phase,
Fig.~\ref{config}(b), which appears to be unstable in 2D due to enhanced 
thermal fluctuations. Another interesting question for the future studies is
whether the partially ordered state can be stabilized in layered
triangular antiferromagnets. The thermal fluctuations are suppressed in this
case by 3D effects, while the mean-field calculations predict
stability of the partially disordered phase up to $J_\perp \sim 0.6 J$.

\end{document}